\documentstyle[12pt,psfig,cite]{article}
\begin{document}
\renewcommand{\thefootnote}{\fnsymbol{footnote}}
\begin{titlepage}
%
%
%
\vspace*{10mm}
\Large
\begin{center}
Random Surfaces in Three-Dimensional Simplicial Gravity
\end{center}
\vspace{5mm}
\normalsize
\begin{center}
H.S.Egawa  \footnote[2]{E-mail address: egawah@theory.kek.jp} 
and 
N.Tsuda,   \footnote[3]{E-mail address: ntsuda@theory.kek.jp}
\end{center}
\begin{center}
$^{\dag}$ Department of Physics, Tokai University\\
Hiratsuka, Kanagawa 259-12, Japan

$^{\ddag}$ Theory Division, Institute of Particle and Nuclear Studies, \\
KEK, High Energy Accelerator Research Organization\\
Tsukuba, Ibaraki 305 , Japan
\end{center}
\vspace{5cm}
\parindent 5cm
\begin{abstract}
A model of simplicial quantum gravity in three dimensions is investigated 
numerically based on the technique of the dynamical triangulation (DT).
We are concerned with the surfaces appearing on boundaries (i.e., sections) 
of three-dimensional DT manifold with $S^{3}$ topology.
A new scaling behavior of genus distributions of boundary surfaces is 
found.
Furthermore, these surfaces are compared with the random surfaces 
generated by the two-dimensional DT method which are well known as a 
correct discretized method of the two-dimensional quantum gravity.
%
\end{abstract}
\end{titlepage}
  
\section{Introduction}
There has been, over the last few years, remarkable progress in the 
quantum theory of two-dimensional gravity.
Two distinct analytic approaches for quantizing two-dimensional gravity 
have been established.
These are recognized as a discretized\cite{matrix} and a 
continuous\cite{Liouville} theory.
The discretized approach, implemented by the matrix model technique, 
exhibits behavior found in the continuous approach, given by Liouville 
field theory, in a continuum limit.
It thus seems to exist strong evidence for the equivalence of the two 
theories in two dimensions. 
Numerical methods based on the matrix model, such as that of dynamical 
triangulation\cite{dt} have drawn much attention as alternative approaches 
to studying non perturbative effects, being also capable of handling those 
cases where analytical theories cannot yet produce meaningful results. 
In the dynamical triangulation method, calculations of the partition function 
are performed by replacing the path integral over the metric to a sum over 
possible triangulations.
Several close studies\cite{fractal} on the structure of two-dimensional DT 
surfaces in quantum gravity were made, and it revealed fractal structures of 
these surfaces. 

However, we feel that the relationship between three-dimensional (also 
four-dimensional) quantum gravity and its discretized model is not yet 
understood.
Over the past few years a considerable number of numerical studies have been 
made on three-dimensional simplicial quantum gravity\cite{3DQG}.
It is generally agreed that the phase transition of three- and four-
dimensional simplicial quantum gravity are both first 
order\cite{3D_1stOD,4D_1stOD}.
The problems are still in controversy.
On the other hand, recent numerical results obtained by the dynamical 
triangulation for three and four dimensions suggest the existences of the 
scaling behavior near to the critical point\cite{3D_Scaling,4D_Scaling,CSP}.
It may be important to note the asymmetric behavior of the order parameter 
in usual phase diagram.
When the coupling strength, $\kappa_{0}$, closes to the critical point from 
the strong coupling side ($\kappa_{0} < \kappa_{0}^{c}$) which corresponds to 
the crumple phase, the transition is smooth.
On the other hand, when $\kappa_{0}$ closes to the critical point from the 
weak coupling side ($\kappa_{0} > \kappa_{0}^{c}$) which corresponds to the 
branched polymer phase, the transition is very rapid.
In ref.\cite{3D_Scaling} it is reported that near the critical point belonging 
to the strong coupling phase the scaling behavior of the mother universe 
exists and also that no mother universe exists in the weak coupling phase 
(i.e., branched polymer phase).
It seems reasonable to suppose that the model makes sense as long as we 
close to the critical point from the strong coupling side.
We actually observe a double peak histogram structure which is a signal 
of the first order phase transition for an appropriate large lattice size.
Therefore, we carefully chose one of peaks belonging to the strong coupling 
phase as an ensemble for our simulations\cite{CSP}.
%

\begin{figure}
\centerline{\psfig{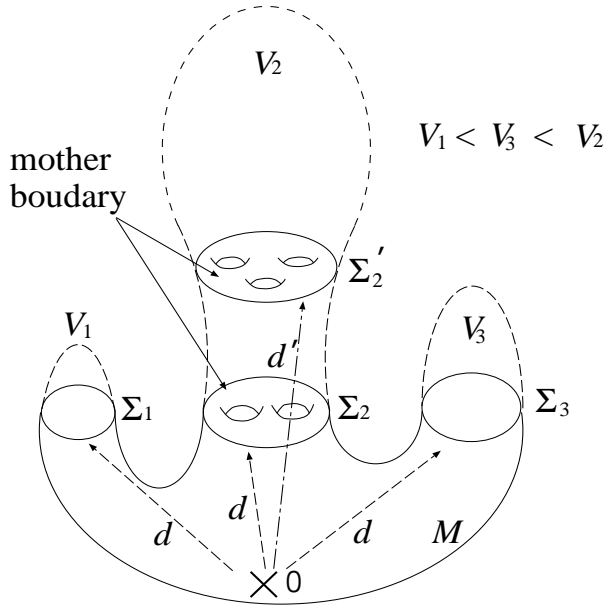}} 
\caption
{
Schematic picture of boundary surfaces ($\Sigma_{1},\Sigma_{2}$ and 
$\Sigma_{3}$) at distance $d$ and a surface ($\Sigma_{2}'$) at distance 
$d'$ in 3D Euclidean space $M$ with $S^{3}$ topology.
The mother boundary surface is defined as a surface ($\Sigma_{2}$) with the 
largest tip volume ($V_{2}$), and the other surfaces are defined as the baby 
boundary.
}
\label{Fig:3D_Boundary}
\end{figure}

%

This paper is intended as two investigations, firstly, the searching a new 
scaling property for the boundary surfaces in three-dimensional DT manifolds 
near the critical point ($\kappa_{0}^{c}$) and, secondary, the relation 
between these boundary surfaces and two dimensional random surfaces 
implemented by the matrix model.
Below we will consider only three-dimensional DT manifolds with $S^{3}$ 
topology.
We can represent a typical configuration in Fig.\ref{Fig:3D_Boundary}.
$M$ denotes a three-dimensional DT manifold with $S^{3}$ topology and 
$\Sigma_{1}$, $\Sigma_{2}$ and $\Sigma_{3}$ denote the boundaries which 
are closed and orientable triangulated surfaces at a distance $d$ from 
an origin ($\times$).
The geodesic distance can be recognized as a usual time parameter.

Our model of three-dimensional Euclidean quantum gravity with boundaries 
may be recognized as a model of a quantum nucleation of the universe in 
$(2+1)$-dimensional gravity.
The nucleation of the universe by a quantum tunneling may be described by 
going out of the Euclidean signature region to the Lorentzian signature 
region in a sense of the semiclassical approximation\cite{FHHMS}.
The nucleation of the universe can be regarded as a topology-changing 
process in the sense that the universe takes a transition from the initial 
state with no boundary to the final state with nontrivial topology.
We can treat the model in a full quantum way which means that we 
sum up all of the fluctuations of the three-dimensional metric $g_{\mu\nu}$ 
and also put no restriction on the boundary surfaces $\Sigma$.
Of course, it is possible to introduce the extrinsic curvature on the 
boundary surface as a physical restriction.
The process of a quantum tunneling requests that all the components of the 
extrinsic curvature vanish (i.e., {\it totally geodesic}).

This paper is organized as follows.
In Sec.2 we briefly review the model of the three-dimensional dynamical 
triangulation.
In Sec.3 all of our numerical results, especially genus distributions 
and coordination number distributions for the mother boundary surfaces, 
are shown.
We summarize our results and discuss some future problems in the final 
section.
%
\section{Model}
It is not yet known how to give a constructive definition of 
three-dimensional quantum gravity.
We start with the Euclidean Einstein-Hilbert action,  
\begin{equation} 
S_{EH} = \int d^3 x \sqrt{g} \left(\Lambda - \frac{1}{G}R \right), 
\end{equation}
where $\Lambda$ is the cosmological constant, and $G$ is Newton's constant 
of gravity.
We use the lattice action of the three-dimensional model with the $S^3$ 
topology, corresponding to the above action, as follows: 
\begin{eqnarray} 
S(\kappa_{0},\kappa_{3}) & = & - \kappa_{0} N_{0} + \kappa_{3} N_{3}  
          \nonumber\\    
  & = & - \frac{2\pi}{G}N_{0} 
        + \left(
           \Lambda' - \frac{1}{G} (2\pi - 6 \mbox{cos}^{-1}(\frac{1}{3}))
          \right) N_{3}, 
\end{eqnarray}
where $N_i$ denotes the total number of $i$-simplexes, and 
$\Lambda' = c\Lambda$; $c$ is the unit volume, and 
$\mbox{cos}^{-1}(\frac{1}{3})$ is the angle between two tetrahedra.
The coupling $\kappa_{0}$ is proportional to the inverse of bare 
Newton's constant, and the coupling $\kappa_{3}$ corresponds 
to a lattice cosmological constant. 

For the dynamical triangulation model of three-dimensional quantum gravity, 
we consider a partition function of the form 
\begin{equation} 
Z(\kappa_{0}, \kappa_{3}) 
= \sum_{T(S^{3})} e^{-S(\kappa_{0}, \kappa_{3})}.
\label{eq:PF} 
\end{equation}
We sum over all simplicial triangulations, $T(S^{3})$, on a 
three-dimensional sphere.
In practice, we must add a small correction term $\Delta S $ to the 
lattice action in order to suppress volume fluctuations. 
The correction term is denoted by  
\begin{equation} 
\Delta S = \delta (N_{3} - N_{3}^{(\mbox{target})})^2,
\end{equation}
where $N_{3}^{(\mbox{target})}$ is the target value of three-simplexes, 
and we use $\delta = 0.0005$ in all our run.
A fine tuning of $\kappa_{0}$ is one of the elaborate works in this model 
because the critical coupling $\kappa_{0}^{c}$ depends on the size of the 
system.
A $\kappa_{0}^{\mbox{equiv}}$ discussed in ref.\cite{3DMCMC} is used as the 
critical value $\kappa_{0}^{c}(N_{3})$ in our simulation.
In ref.\cite{3DMCMC} the parameter $\kappa_{0}$ is tuned to 
$\kappa_{0}^{\mbox{equiv}}$ using the multicanonical Monte Carlo method such 
that the heights of double peaks of $N_{0}$ become the same.
%

\section{Measurements}
We now define the intrinsic geometry using the concept of a geodesic 
distance as a minimum length (i.e., minimum step) in the dual lattice 
between two tetrahedrons in a three-dimensional DT space.
Suppose a tree-dimensional ball (3-ball) which is covered within $d$ 
steps from a reference 3-simplex in the three-dimensional manifold with 
$S^{3}$ topology. 
Naively, the 3-ball has a boundary with spherical topology ($S^{2}$).
However, because of the branching of the DT space, the boundary is not 
always simply-connected, and there usually appear many boundaries which 
consist of closed and orientable two-dimensional surfaces with any 
topology and nontrivial structures such as links or knots\footnote{In the 
strict sense links or knots are constructed by loops. 
In our case the loop is a fat loop.}.
These two-dimensional boundary surfaces are equivalent to two-dimensional 
randomly triangulated surfaces with corresponding genuses.
In order to discuss the scaling properties of these surfaces, the genus 
distributions of these surfaces are measured. 
We should notice that there appear many boundaries with distance $d$ 
(see Fig.\ref{Fig:3D_Boundary}).
The boundaries are divided into two classes: one is a baby boundary and the 
other is a mother one.
The boundaries with small sizes ($\sim {\cal O}(1)$) are called by a baby 
one.
The baby boundaries are originated from the small fluctuations of the 
three-dimensional Euclidean spaces.
We thus think that these surfaces are non-universal objects 
and then suffer from large finite size effects.
In this section, we shall concentrate only on the mother boundary.

Here, we give a precise definition of ``baby'' and ``mother'' universes in 
Fig.\ref{Fig:3D_Boundary}.
The mother universe is defined as a boundary surface ($\Sigma_{2}$) with the 
largest volume ($V_{2}$), and the other surfaces are defined as baby 
universes.
%
\begin{figure}
\centerline{\psfig{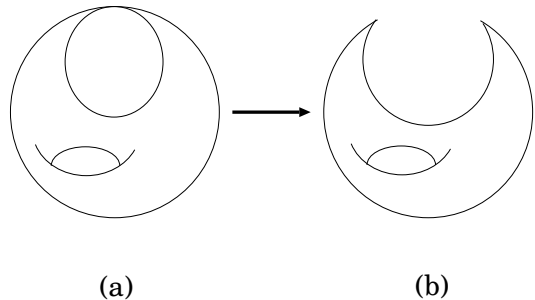}} 
\caption
{
Measurement of a genus of boundary surface.
(a) denotes an example of a irregular configuration.
The boundary surfaces in general may have many singular vertices and links.
(b) denotes a regular configuration.
After we change (a) into (b), we measure the genus of (b).
}
\label{Fig:True_Genus}
\end{figure}

\begin{figure}
\centerline{\psfig{file=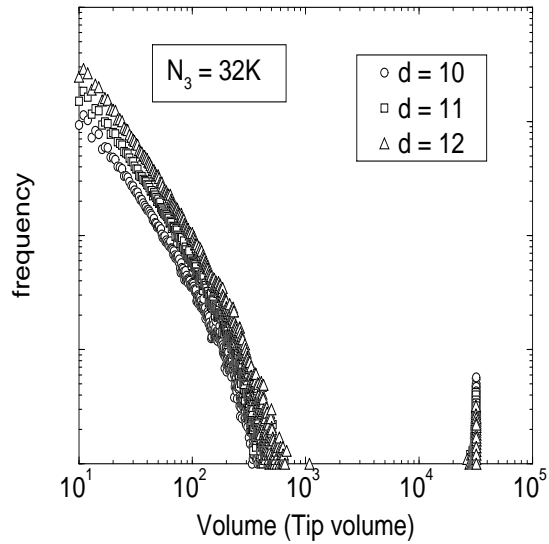,height=8cm,width=8cm}} 
\caption
{
Typical distributions of the tip volume distributions for all the boundaries 
with $N_{3}=32K$ and $d=10,11$ and $12$. Discrepancies shown in a small part 
of the tip volume ($10 \sim 100$) are due to the finite 
size effects.
}
\label{Fig:Tip_Vol_32K}
\end{figure}

%
The problem is whether we can distinguish these two kinds of universes 
(baby and mother boundary surfaces) in our simulations.
The answer is yes.
In Fig.\ref{Fig:Tip_Vol_32K} we plot the tip volumes 
(correspond to $V_{1}$, $V_{2}$ and $V_{3}$ in Fig.1) distributions near to 
the critical point ($\kappa_{0} = 4.195$ and $\kappa_{3} = 2.222$) with 
appropriate geodesic distances when $N_{3}=32$.
This figure tells us that there is one mother boundary whose tip volume is 
prominent and the others belong to the class of the baby universe.
At any rate, we can easily distinguish the mother boundary surface from 
the others by the measurement of the tip volume.

In ref.\cite{3D_Scaling} it is reported that the surface-area-distributions 
(SAD) of the mother universe show the scaling behavior near to the critical 
point but the baby universe does not show a scaling behaviour at all.
The important point to note is that in ref.\cite{3D_Scaling} the measurements 
in the critical region were done in the strong coupling phase.
Therefore, we focus on the mother universe near to the critical point taking 
the limit from the strong coupling phase following subsections $3.1$ and $3.3$.
On the other hand, it is known that in the weak coupling limit the 
three-dimensional DT manifold becomes a branched polymer, and its boundary 
surfaces show no scaling property\cite{3D_Scaling}.
Therefore, we cannot consider the genus distributions\footnote{We obtain only 
spherical boundary surfaces at any geodesic distances in this phase.} of 
boundary surfaces in the weak coupling phase.

\subsection{Genus distributions of boundary surfaces near to the critical 
point}
%

%
\begin{figure}
\vspace{8mm}
\hspace*{-9mm}
\centerline{\psfig{file=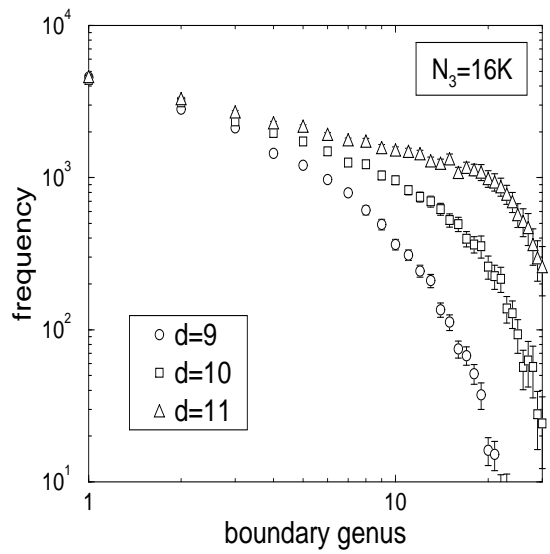,height=8cm,width=8cm}} 
\caption
{
Genus distributions of mother boundary at the critical point for various 
distances with log-log scales.
$N_{3}=16K$ ($\kappa_{0}^{c}=4.090$ and $\kappa_{3}=2.200$). 
$d=9,10$ and $11$.
}
\label{Fig:BG_Critical_V16K}
\end{figure}

%
\begin{figure}
\vspace{8mm}
\hspace*{-9mm}
\centerline{\psfig{file=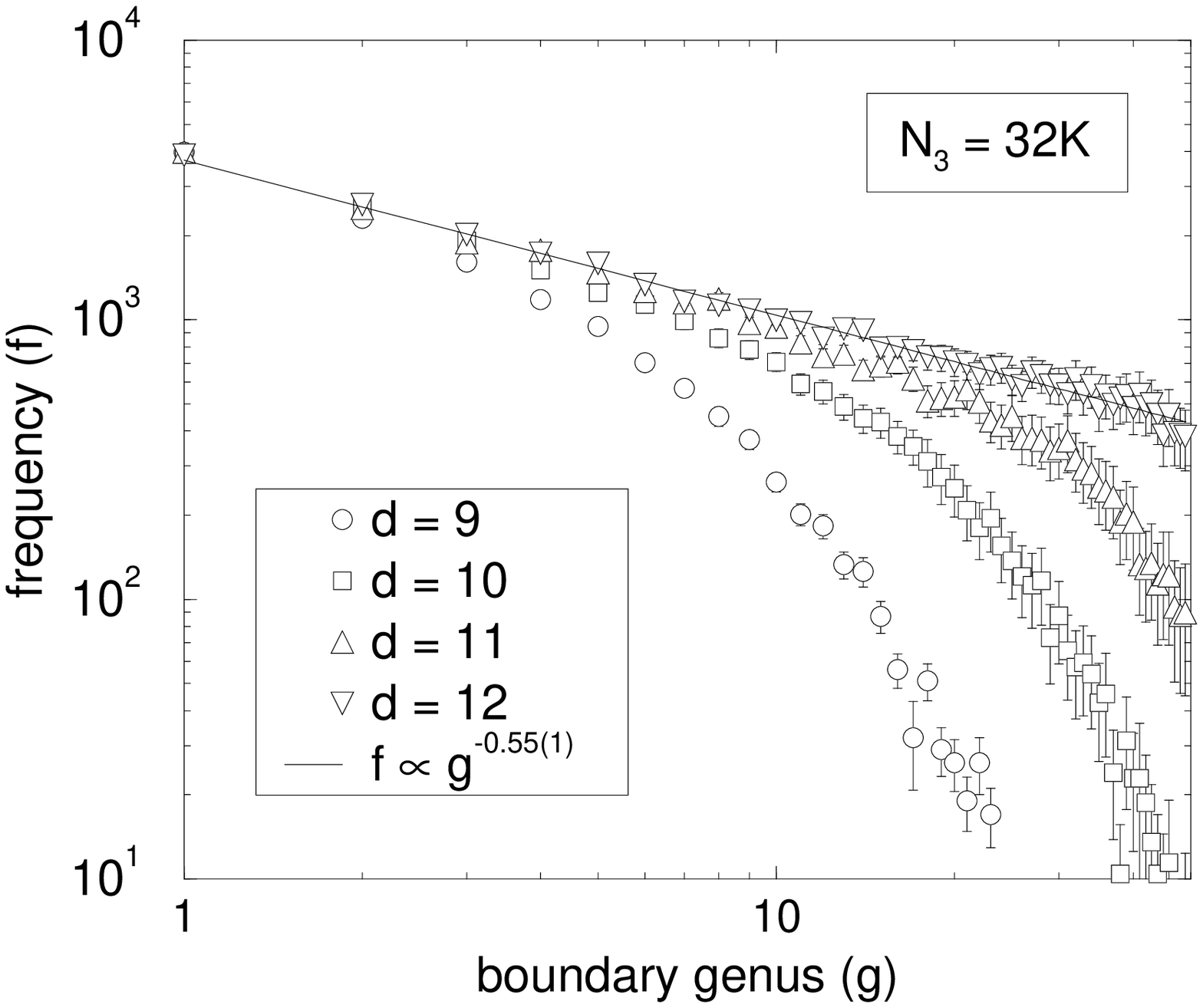,height=8cm,width=8cm}} 
\caption
{
Genus distributions of mother boundary at the critical point for various 
distances with log-log scales. 
$N_{3}=32K$ ($\kappa_{0}^{c}=4.195$ and $\kappa_{3}=2.222$). 
$d=9,10,11$ and $12$.
Solid line is drawn by the least squares of the data with $d=12$ with 
$1\leq g \leq 50$ as a fitting range.
}
\label{Fig:BG_Critical_V32K}
\end{figure}

%
An interesting observable for three-dimensional DT manifolds is a genus 
distribution of boundary surfaces with various geodesic distances.
The genus of the mother boundary surfaces is, in fact, naively expected 
to scale because the genus is a dimensionless quantity.
The boundary surfaces in general have very complicated and irregular 
structures.
We illustrate an irregular surface as a simple example: ($a$) in 
Fig.\ref{Fig:True_Genus}.
In order to calculate the genus of these boundary surfaces we must 
dispose these irregular configurations.
We thus detach irregular sub-simplexes (vertices or links) like in 
Fig.\ref{Fig:True_Genus} ($(a) \to (b)$) 
\footnote{
This deformation ($(a) \to (b)$) is recognized as the infinitesimal deformation 
$d \; \to \; d - \varepsilon$. Since $\varepsilon$ can be arbitrary small, we 
obtain the same distributions of the boundary genus as at distance $d$ in the 
limit $\varepsilon \to 0$.
}

The small mother boundary surfaces will suffer a large finite-size effect, 
so that almost all these surfaces become spherical surfaces.
Therefore, we must introduce the lower limit ($V_{cut}$) of the volume for 
the spherical mother surface in order to avoid the large finite-size effects.
In this subsection we ignore spherical surfaces for simplicity.
In next subsection $3.2$, we actually introduce the lower limit ($V_{cut}$) 
in order to measure coordination number distributions.
Figs.\ref{Fig:BG_Critical_V16K} and \ref{Fig:BG_Critical_V32K} shows genus 
distributions of the mother boundary surfaces $P(g)$ with various distances 
and with $N_{3}=16K$ and $32K$
\footnote{
In Figs.\ref{Fig:BG_Critical_V16K} and \ref{Fig:BG_Critical_V32K}  
we prepared about $400 (N_{3}=16K)$ and $100 (N_{3}=32K)$ configurations 
using the Monte Carlo method. 
We took $160$ start points (simplices) for $N_{3}=16K$ and and $320$ for 
$N_{3}=32K$ par a configuration.
}.
We naively expect that the genus distributions of the boundary 
surfaces will show the scaling properties and, in fact, we find scaling 
behavior ,
\begin{equation}
P(g) \sim g^{-\alpha},
\end{equation}
where $g$ is the genus and $\alpha$ is $0.55(1)$.
$\alpha$ is obtained by the least squares from the data in 
Fig.\ref{Fig:BG_Critical_V32K} with $N_{3} = 32K$ and with 
$1\leq g \leq 50$ as a fitting range.
It reveals that the scaling property of the genus distributions becomes 
the more clear the bigger the size of boundary becomes.
Then, this scaling property will remains after the thermal limit $N_{3} 
\to \infty$.
If $\kappa_{0}$ is out of the critical point this scaling relation 
disappears.
%

\subsection{Genus distributions in the strong coupling phase}
\begin{figure}
\centerline{\psfig{file=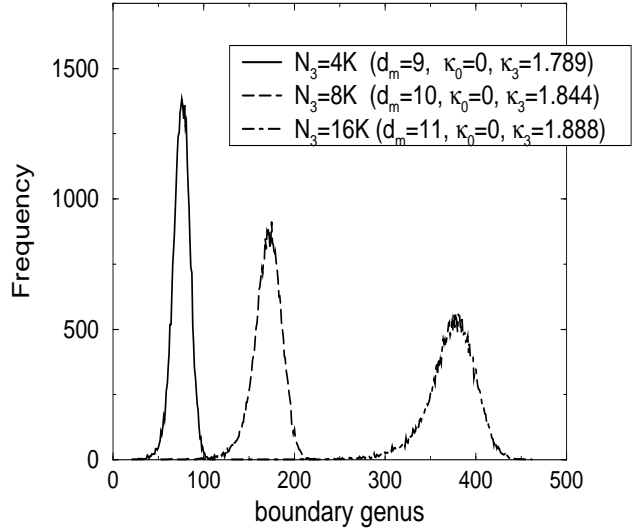,height=8cm,width=8cm}} 
\caption
{
Genus distributions of the mother boundary surface in the strong coupling 
phase for various sizes.
}
\label{Fig:Boundary_Genus_Strong}
\end{figure}

We also measure the genus distributions of the mother boundary surface in 
the strong coupling limit (i.e., $\kappa_{0}=0$) with various $N_{3}$ and 
distances.
Fig.\ref{Fig:Boundary_Genus_Strong} shows the distributions of the mother 
boundary surfaces with three different volumes ($N_{3}=4K,8K$ and $16K$)
\footnote{In Fig.\ref{Fig:Boundary_Genus_Strong} we have prepared 
$800(N_{3}=4K),400(N_{3}=8K)$ and $200(N_{3}=16K)$ configurations and 
obtain $32,000$ genus data points evenly for each volume.}, and it reveals 
that these distributions seem to the Gaussian.
In this measurement we select a distance ($d_{m}$) at which the peak value 
of each distribution becomes maximum, and actually in 
Fig.\ref{Fig:Boundary_Genus_Strong} we use the values $d_{m}=9,10$ and $11$ 
for $N_{3}=4K,8K$ and $16K$, respectively.
The peak value ($g_{p}$) of each distribution becomes the larger the bigger 
the size of boundary surface becomes.
We know that the manifold becomes crumple in this phase, and the fractal 
dimension seems to diverge.
We, therefore, reasonably conclude that $g_{p}$ diverges when $N_{3}$ goes 
to infinity and that the manifold becomes crumple boundlessly in the strong 
coupling limit.
%

%
\subsection{Coordination number distributions of boundary surfaces}
We must look more carefully into these boundary mfds.
In the last few years, several articles have been devoted to the study of 
boundary mfds.
In two dimensions, it is revealed by ref.\cite{IK} that the dynamics of the 
string world sheet (random surfaces) can be described by the time 
\footnote{i.e., geodesic distances} evolution of boundary loops.
Furthermore, the work\cite{SS} is also based on the idea that the functional 
integral for $3$ and $4$D quantum gravity can be represented as a superposition 
of less complicated theory of random surfaces
\footnote{In this case a definition of a time direction is different 
from our definition of a time slice, i.e., geodesic distance.}.
It is precisely on such grounds that we claim that the higher 
dimensional complicated theory of quantum gravity can be reduced to the 
lower dimensional quantum gravity.

\begin{figure}
\centerline{\psfig{file=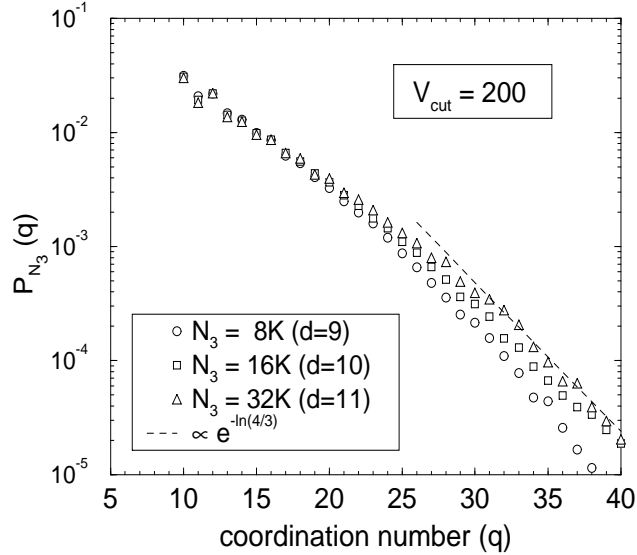,height=8cm,width=9cm}} 
\caption
{
Normalized coordination number distributions for the mother boundary 
universe near to the critical point with $N_{3}=8K (\kappa_{0} = 3.950$ and 
$\kappa_{3} = 2.165),16K (\kappa_{0} = 4.090$ and $\kappa_{3} = 2.200)$ and 
$32K (\kappa_{0} = 4.195$ and $\kappa_{3} = 2.222)$ with linear-log scales 
and with $V_{cut} = 200$.
We have measured the distributions only for the mother boundary surfaces with 
spherical ($S^{2}$) topology.
Broken line is a guide line and denotes the asymptotic scaling relation: 
$P_{N_{3}}(q) \propto e^{-\ln{\frac{4}{3}} q}$.
}
\label{Fig:Boundary_V.O_Comp}
\end{figure}

The question is that the closed and orientable two-dimensional boundary 
surfaces mentioned above can be recognized as the two-dimensional random 
surfaces described by the matrix model or the Liouville field theory.
In this subsection, we focus on the distributions of the coordination 
numbers.
The distributions of the coordination numbers have been calculated in 
ref.\cite{BIPZ}.
The probability distribution $P_{N}(q)$ can be extracted from the Green 
functions of the $\phi^{3}$ planar theory with spherical ($S^{2}$) topology 
without self-energy and tadpole graph in dual lattice, 
\begin{equation}
P_{N \to \infty}(q) = 16 (\frac{3}{16})^{q} \frac{(q-2)(2q-2)!}
{q!(q-1)!} \stackrel{q \to \infty}{\longrightarrow} e^{-\ln{\frac{4}{3}}q},
\label{eq:coordination}
\end{equation}
where N is a number of 2-simplexes and q is the coordination number.
The probability distribution $P_{N}(q)$ with small $q$ region is strongly 
dependent on the local lattice structures.
The value ($\ln{\frac{4}{3}}$) guarantees the randomness of the lattice 
(see Appendix A).
In the preceding subsection we pointed out that spherical mother boundary 
surfaces whose volumes were greater than $V_{cut}$ had to be considered.
Fig.\ref{Fig:Boundary_V.O_Comp} shows the (correct-normalized) coordination 
number distributions of the spherical mother boundary surface for various 
volumes with $V_{cut} = 200$.
In Fig.\ref{Fig:Boundary_V.O_Comp} we also show the result obtained by 
the theoretical calculation eq.(\ref{eq:coordination}) (broken line).
%

%
The data of the coordination number distributions of the spherical mother 
boundary surfaces become consistent with the distributions 
$P_{N}(q) \propto e^{-\ln{\frac{4}{3}} q}$ for large enough $q$.
On the other hand, there are a few discrepancies between the two-dimensional 
theoretical curve and the small $q$ ($q < \sim 25$) distributions in our 
simulation sizes.
We cannot say from only these data whether the mother boundary surfaces with 
$S^{2}$ topology are equivalent to the random surfaces implemented by the 
matrix model or not.
Further discussion will be presented in the next section.

\section{Summary and Discussion}
We investigate the boundary surfaces of three-dimensional DT manifolds with 
$S^{3}$ topology.
It have been naively expected that the genus distributions of boundary 
surfaces will show the scaling properties near the critical point and, in 
fact, we find scaling behavior, $P(g) \sim g^{-\alpha},$ where $\alpha$ is 
about $0.5$ in our simulation sizes.
On these grounds we may conclude that the thermodynamic limit ($N_{3} \to 
\infty$) can be taken in this scaling region.

At the strong coupling limit ($\kappa_{0}=0$) the genus distribution of mother 
universe seems to Gaussian distribution, and it is non-universal in the sense 
that the peak value of genus $g_{p}$ diverges as $N_{3} \to \infty$.

On the other hand, in the weak coupling region, all of the size of the 
boundary surfaces are very small, $\sim {\cal O}(1)$, and then these genuses 
are strongly restricted to zero (i.e., sphere).
As a result, there is no scaling behaviour of the genus distributions in the 
weak coupling phase.

Our numerical results show that the coordination number distributions of the 
spherical mother boundary surfaces of three-dimensional DT manifolds are 
consistent with the theoretical prediction in the large $q$ region.
When these boundary surfaces can be recognized as surfaces of the matrix 
model or the Liouville theory, three-dimensional DT manifolds can be 
reconstructed by the direct products of $\Sigma$ (two-dimensional DT 
surfaces) and $d$ (geodesic distance).
In order to confirm the equivalence between the boundary surfaces in 
three-dimensions and random surfaces in two-dimensions the string susceptibility 
$\gamma_{str}$ of the boundary surfaces or the loop-length-distributions (LLD) of 
the boundary surfaces must be measured and compared with theoretical predictions.
We have obtained some preliminary results of the string susceptibility 
exponents of the boundary surfaces, and a more complete numerical analysis 
will be published elsewhere.

We will be able to apply our numerical analysis to other physical process 
such as the topology changing of surfaces of $(2+1)$-dimensional gravity.
In this process we substitute the geodesic distance ($d$) for the time, and 
the two-dimensional surfaces are recognized as the boundary of 
three-dimensional Euclidean manifold.

Furthermore, the boundary surfaces of three dimensions in general have other 
nontrivial structures such as Hopf's link and the nontrivial knot in the 
knot theory.
Up to now, it is difficult to extract the informations for the extrinsic 
geometry, i.e., how to embed the boundary surface into {\bf R}$^{3}$ because 
we have only intrinsic geometries of DT manifold.
To know a mechanism of the formation of nontrivial links or knots is one of 
challenging subjects in the statistical mechanics.

\newpage

\begin{center}
{\Large Appendix A}
\end{center}
We can obtain the asymptotic coordination number distribution: 
eq.(\ref{eq:coordination}) only assuming the Poisson distribution function as 
follows,
$$
P(q) \sim e^{-\sigma q}.
\eqno{(A.1)}
$$
By using this and the lower limit of $q > 3$ we can calculate an average 
coordination number $<q>$
$$
<q> = \frac{\sum q e^{-\sigma q}}{\sum e^{-\sigma q}} = 
-\frac{d}{d\sigma}\ln{(\sum_{q>3} e^{-\sigma q})} \stackrel{N_{0} \to \infty}
{\longrightarrow} 3 + \frac{e^{-\sigma}}
{1 - e^{-\sigma}}
\eqno{(A.2)}
$$
On the other hand, $<q>$ is trivial in two dimensional DT mfd,
$$
<q> = \frac{2N_{1}}{N_{0}} = 6 - \frac{\chi}{N_{0}} \approx 6, 
\eqno{(A.3)}
$$
where $\chi$ is the Euler number.
From eqs.(A.2) and (A.3) we obtain $\sigma = \ln{\frac{4}{3}}$ \cite{NTD}.

\begin{center}
{\Large Acknowledgements}
\end{center}
We are grateful to H.Kawai, T.Yukawa, H.Hagura and T.Izubuchi for useful 
discussions and comments.
One of the authors (N.T.) is supported by Research Fellowships of the Japan 
Society for the Promotion of Science for Young Scientists.

\newpage


\end{document}